\documentclass[oneside]{elsart}
\usepackage{epsfig}
\usepackage{caption2}
\usepackage{graphicx}
\newcommand{\ks}{\mbox{$K_{S}$}}
\newcommand{\kl}{\mbox{$K_{L}$}}

\newcommand{\aff}[2]{Dipartimento di Fisica dell'Universit\`a #1 e Sezione INFN, #2, Italy.}
\newcommand{\affd}[1]{Dipartimento di Fisica dell'Universit\`a e Sezione INFN, #1, Italy.}
\begin{document}
\begin{frontmatter}
\pagenumbering{arabic}
\pagestyle{plain}
\title{ \bf{
Branching ratio measurement of $K_S\rightarrow \gamma \gamma$ 
decay using a pure $\ks$ beam} in the KLOE detector}
\collab{The KLOE Collaboration}
\author[Na]{F.~Ambrosino},
\author[Frascati]{A.~Antonelli},
\author[Frascati]{M.~Antonelli},
\author[Frascati]{F.~Archilli},
\author[Roma3]{C.~Bacci},
\author[Karlsruhe]{P.~Beltrame},
\author[Frascati]{G.~Bencivenni},
\author[Frascati]{S.~Bertolucci},
\author[Roma1]{C.~Bini},
\author[Frascati]{C.~Bloise},
\author[Roma3]{S.~Bocchetta},
\author[Roma1]{V.~Bocci},
\author[Frascati]{F.~Bossi},
\author[Roma3]{P.~Branchini},
\author[Roma1]{R.~Caloi},
\author[Frascati]{P.~Campana},
\author[Frascati]{G.~Capon},
\author[Na]{T.~Capussela},
\author[Roma3]{F.~Ceradini},
\author[Frascati]{S.~Chi},
\author[Na]{G.~Chiefari},
\author[Frascati]{P.~Ciambrone},
\author[Frascati]{E.~De~Lucia},
\author[Roma1]{A.~De~Santis},
\author[Frascati]{P.~De~Simone},
\author[Roma1]{G.~De~Zorzi},
\author[Karlsruhe]{A.~Denig},
\author[Roma1]{A.~Di~Domenico},
\author[Na]{C.~Di~Donato},
\author[Pisa]{S.~Di~Falco},
\author[Roma3]{B.~Di~Micco},
\author[Na]{A.~Doria},
\author[Frascati]{M.~Dreucci},
\author[Frascati]{G.~Felici},
\author[Frascati]{A.~Ferrari},
\author[Frascati]{M.~L.~Ferrer},
\author[Frascati]{G.~Finocchiaro},
\author[Roma1]{S.~Fiore},
\author[Frascati]{C.~Forti},
\author[Roma1]{P.~Franzini},
\author[Frascati]{C.~Gatti},
\author[Roma1]{P.~Gauzzi},
\author[Frascati]{S.~Giovannella},
\author[Lecce]{E.~Gorini},
\author[Roma3]{E.~Graziani},
\author[Pisa]{M.~Incagli},
\author[Karlsruhe]{W.~Kluge},
\author[Moscow]{V.~Kulikov},
\author[Roma1]{F.~Lacava},
\author[Frascati]{G.~Lanfranchi},
\author[Frascati,StonyBrook]{J.~Lee-Franzini},
\author[Karlsruhe]{D.~Leone},
\author[Frascati]{M.~Martini},
\author[Na]{P.~Massarotti},
\author[Frascati]{W.~Mei},
\author[Na]{S.~Meola},
\author[Frascati]{S.~Miscetti},
\author[Frascati]{M.~Moulson},
\author[Frascati]{S.~M\"uller},
\author[Frascati]{F.~Murtas},
\author[Na]{M.~Napolitano},
\author[Roma3]{F.~Nguyen},
\author[Frascati]{M.~Palutan},
\author[Roma1]{E.~Pasqualucci},
\author[Roma3]{A.~Passeri},
\author[Frascati,Energ]{V.~Patera},
\author[Na]{F.~Perfetto},
\author[Lecce]{M.~Primavera},
\author[Frascati]{P.~Santangelo},
\author[Na]{G.~Saracino},
\author[Frascati]{B.~Sciascia},
\author[Frascati,Energ]{A.~Sciubba},
\author[Pisa]{F.~Scuri},
\author[Frascati]{I.~Sfiligoi},
\author[Frascati]{T.~Spadaro},
\author[Roma1]{M.~Testa},
\author[Roma3]{L.~Tortora},
\author[Roma1]{P.~Valente},
\author[Karlsruhe]{B.~Valeriani},
\author[Frascati]{G.~Venanzoni},
\author[Frascati]{R.Versaci},
\author[Frascati,Beijing]{G.~Xu}

\address[Frascati]{Laboratori Nazionali di Frascati dell'INFN, 
Frascati, Italy.}
\address[Karlsruhe]{Institut f\"ur Experimentelle Kernphysik, 
Universit\"at Karlsruhe, Germany.}
\address[Lecce]{\affd{Lecce}}
\address[Na]{Dipartimento di Scienze Fisiche dell'Universit\`a 
``Federico II'' e Sezione INFN,
Napoli, Italy}
\address[Pisa]{\affd{Pisa}}
\address[Energ]{Dipartimento di Energetica dell'Universit\`a 
``La Sapienza'', Roma, Italy.}
\address[Roma1]{\aff{``La Sapienza''}{Roma}}
\address[Roma3]{\aff{``Roma Tre''}{Roma}}
\address[StonyBrook]{Physics Department, State University of New 
York at Stony Brook, USA.}
\address[Beijing]{Permanent address: Institute of High Energy 
Physics of Academica Sinica,  Beijing, China.}
\address[Moscow]{Permanent address: Institute for Theoretical 
and Experimental Physics, Moscow, Russia.}

\begin{abstract}
We have analyzed 1.62 fb$^{-1}$ of $e^{+}e^{-}$ collisions 
at a center of mass energy  $\sim M_{\phi}$ collected
by the KLOE experiment at DA$\Phi$NE. This sample corresponds 
to a production of $\sim$ 1.7 billion of $\ks$ $\kl$ pairs
which allowed us to search for the rare $K_S\to \gamma\gamma$
decay. $K_S$ are tagged by the   $K_L$ interaction in the calorimeter
and the signal is searched for by requiring two additional prompt 
photons. Strong kinematic requirements reduce
the initial 0.5$\times 10^6$ events to 2300
candidates from which we extract a signal of 600 $\pm$ 35 events. 
By normalizing to the $\ks\ \to 2 \pi^0$ decays
counted in the same sample, the measured value of 
BR($\ks \to \gamma\gamma$) is
(2.27 $\pm 0.13(stat.) ^{+0.03} _{-0.04} (syst.)) \times 10^{-6}$,
in agreement with $O(P^4)$ Chiral Perturbation Theory predictions.
\end{abstract}

\end{frontmatter}
\baselineskip=14pt
\section{Introduction}
A precise measurement of the $K_S\to \gamma\gamma$ decay rate is an
important test of Chiral Perturbation Theory ($\chi PT$) predictions.
The decay amplitude of $K_S\to \gamma\gamma$ has been evaluated 
at the leading order of $\chi PT$~\cite{ambrosio},  $O(P^4)$, 
providing the estimate of the corresponding branching ratio, 
$BR(KS\rightarrow\gamma\gamma)= 2.1\times 10^{-6}$, with a few percent 
precision. This result is in agreement with the experimental 
measurement of NA31, that obtained 
$BR(K_S\rightarrow\gamma\gamma)=(2.4\pm 0.9)\times 10^{-6}$ \cite{na31}.
The last precise determination of this $BR(K_S\rightarrow\gamma\gamma)$ of $2.71\times 
10^{-6}$, with a total uncertainty below 3\%, comes from
NA48 \cite{na48}. The last mentioned result differs from $\chi PT$ $O(P^4)$ 
prediction of about 30\%, indicating possible contributions from higher order corrections.

In this paper, we show our search based on a data sample 
of 1.62 fb$^{-1}$ of $e^+ e^-$ collisions collected 
with the KLOE detector~\cite{kloe1}~-~\cite{kloe4} at 
DA$\Phi$NE~\cite{dafne}, the Frascati $\phi$-factory.
DA$\Phi$NE is an $e^+e^-$ collider which operates at a 
center of mass energy, $W$, of $\sim 1020$~MeV,
the mass of the $\phi$-meson. Equal-energy positron and
electron beams collide at an angle of ($\pi$ - 25 mrad) producing
$\phi$-mesons nearly at rest. $\phi$-mesons decay 34\% of
the time into nearly collinear $K^{0}\overline K^0$ pairs.
Since $J^{PC}(\phi)=1^{--}$, these pairs are in an
antisymmetric state so the final state is always
$K_SK_L$. All these imply that detection of
a $K_L$ guarantees the presence of a $K_S$ of given momentum
and direction. KLOE takes advantage of this to identify $K_S$-mesons
independent of the decay mode. We refer to it as $K_S$ {\em tagging}.

The data sample analyzed corresponds to a production of $\sim$
1.7 billions of $K_S K_L$  pairs. In the analysis an equivalent statistics of 
simulated events for the background was produced, as well as a sample with simulation of 
the signal with a factor $\sim$ 15 larger.
Using these samples allow us to reach a statistical error of 5.6\% on the 
signal. While this  accuracy is statistically inferior to the most 
precise NA48 result, this new measurement, having completely different  background composition and origin of systematics, as well as from a pure $K_S$ beam, can help 
to clarify  whether $O(P^6)$ contribution are present.

\section{The KLOE detector}

The KLOE detector consists of a large cylindrical drift 
chamber, DC~\cite{kloe1}, of 4~m diameter and 3.3~m length
with an helium-based gas mixture, surrounded by a 
lead-scintillating fiber electromagnetic calorimeter, 
EMC~\cite{kloe2}. A superconducting coil around 
the EMC provides a 0.52 T field. Permanent quadrupoles
for beam focusing are inside the apparatus and are
surrounded by two compact tile calorimeters with veto 
purposes, QCAL~\cite{kloe3}. In this analysis, only the
calorimeter system is used.

The calorimeter is
divided into a barrel and two endcaps covering 98\% of the solid angle. 
The modules are read out at both ends by 
photomultipliers with a $\sim$ 4.4$\times$4.4~cm$^2$
readout granularity, for a total of 2440 cells. Both amplitude and time signals are time signals collected at  the two ends. The amplitude gives a measure of the energy deposited in the modules and the time signal yields both the arrival time of 
particles and the position in three dimensions  of the energy deposits.
Cells close in time and space are grouped into a "calorimeter cluster". 
The "cluster energy" $E$ is the sum of the cell energies. 
The cluster time $T$ and position $\vec{R}$ 
are energy-weighted averages. 
Energy and time resolutions are $\sigma_E/E = 
5.7\%/\sqrt{E\ {\rm(GeV)}}$ and  
$\sigma_t = 57\ {\rm ps}/\sqrt{E\ {\rm(GeV)}}
\oplus50\ {\rm ps}$ respectively.

The QCAL detector comprises two tile calorimeters of $\sim 5 X^0$ placed
thickness close to the IP,  surrounding the focusing quadrupoles.
Each calorimeter consists of a sampling structure of lead and scintillator
tiles arranged in 16 azimuthal sectors. The readout is done 
via wavelength shifter (WLS) fibers coupled to mesh 
photomultipliers. The special arrangement of WLS fibers allows also the measurement of the
longitudinal coordinate by time differences. The tiles are
assembled to maximize efficiency for photons coming from the $K_L$
decays, yet retaining a high efficiency also for photons
coming from the IP. The QCAL solid angle coverage is 
$0.94 < |cos \theta|< 0.99$.

The standard KLOE trigger \cite{kloe4} uses calorimeter and chamber 
information. For this analysis only the calorimeter signals are relevant.
Two energy deposits with $E>50$ MeV for the barrel and  $E>150$ MeV for the 
endcaps are required. Identification and rejection of cosmic-ray events
are also done at the trigger level.

\section{Search of $K_S\to\gamma\gamma$ with a pure $\ks$ beam}
 
\subsection{$K_S$ tagging}
At the center of mass energy of $M_{\phi}$, 
the mean decay lengths of the $K_S$ and $K_L$ are
 $\lambda_S \sim 0.6$ cm and $\lambda_L \sim 340$ cm
respectively.
About 50 \% of $K_L$'s reach the calorimeter before 
decaying.
$\ks$'s  are tagged with high efficiency ($\sim 30\%$) 
by identifying a $\kl$ interaction, which we call "$K_L$-crash". This "$K_L$-crash" has a very distinctive signature in the calorimeter, given by a late 
($\beta_K = 0.2)$ high-energy cluster un-associated to any track. The "$K_L$-crash"
provides a clean $K_S$ tag. In this analysis the fake-tag contribution is essentially negligible. The average value of the center of mass energy  $W$, is obtained with a precision of 30 keV for each 100 nb$^{-1}$ running period, by reconstructing large
angle Bhabha scattering events. The value of $W$ and the "$K_L$-crash" cluster position
allow us to establish, for each event, the trajectory of the $K_S$ 
with an angular resolution of 1$^{\circ}$ and a momentum
resolution of $\sim$ 2 MeV.  

We use for normalization the measurement of the dominant neutral decay mode of
the $\ks \to 2 \pi^0$, always tagged by "$K_L$-crash".

In the analyzed sample, we have $\sim$ 480 $\times 10^6$ 
$\ks$ tagged events. Using the most precise value
of $BR{KS}\rightarrow{\gamma\gamma}$, we expect $\sim 1300$
$K_S \to \gamma \gamma$ events  to be produced and tagged. 
The advantage of the present measurement is that, by tagging, 
we can completely neglect the $K_L\to\gamma\gamma$ background, 
which is the major contamination in NA48 analysis.

\begin{figure}[!t]
\begin{center}
\epsfig{width=.9\textwidth,file=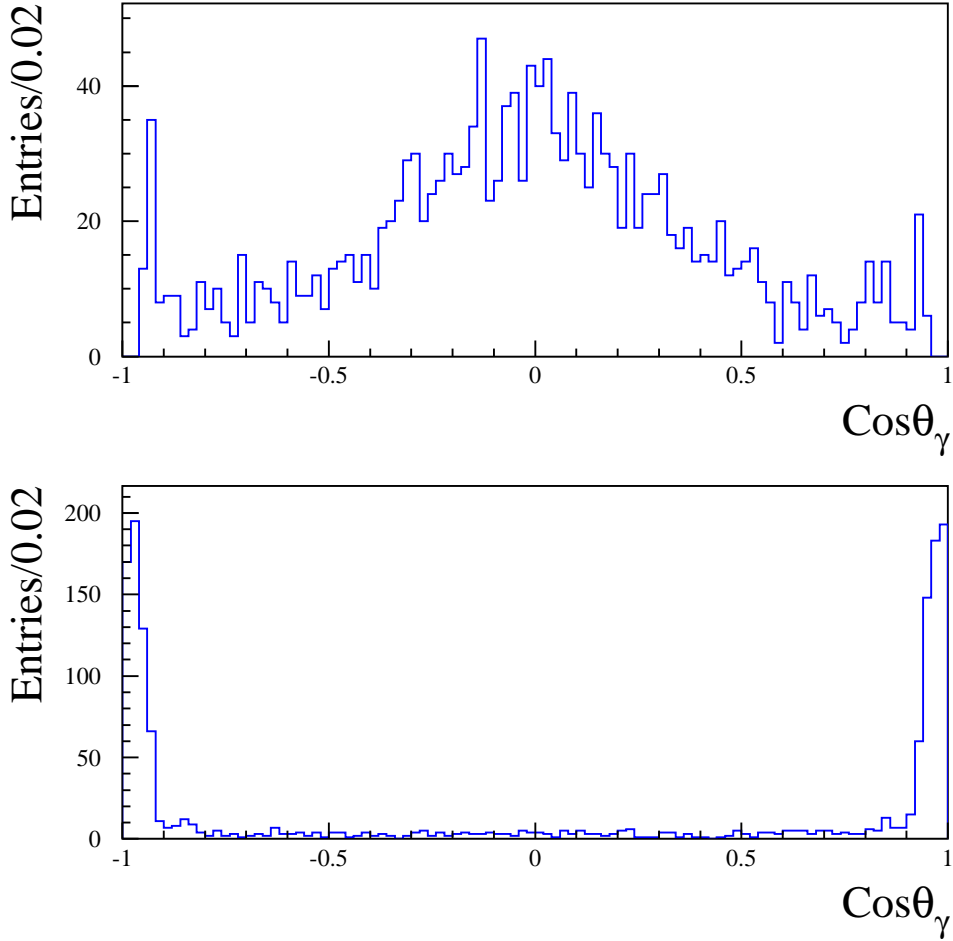}
\end{center} 
\caption{Angular distribution of the generated photons for
$K_S \to 2 \pi^0$ events after "$K_L$-crash" tagging: 
photons reconstructed by the EMC calorimeter (top),
lost photons (bottom).}
\label{QCALMC}
\end{figure}

\subsection{Simulation of background and signal}

The main expected background in this search are 
$K_S\rightarrow 2\pi^0$ events with two lost photons.
Such losses can be due to photons 1) either out of
acceptance, not reconstructed by the calorimeter or 2) merged together.
For the simulation of the background Monte Carlo, MC, we use
a production of $\phi \to K_S K_L$ decays corresponding to 
an equivalent statistics of $\sim $ 1.1 fb$^{-1}$.
For the MC signal we used instead, a production equivalent to
$\sim$ 18 fb$^{-1}$.

The photon properties in the simulation (resolutions and detection-efficiency)
have been tuned with data using a large sample of tagged photons 
in $\phi \to \pi^+\pi^-\pi^0$  events \cite{NIM}
selected using only drift chamber information.
The presence of additional clusters in the 
events due to accidental overlap with machine background, 
{\em dafbkg}, clusters has been taken into account 
by inserting these events in the MC at hit level in 
the simulation. The {\em dafbkg} insertion includes its
rate dependence along the running period, and all other  basic running conditions, such
as beam parameters ($W,\overline P_{\phi},\overline X_{\phi}$) which vary run by run.
The interaction of the $K_L$ in the calorimeter is also properly
simulated. 

\subsection{Event preselection}
\label{accsel}

After tagging, we select events by counting the number of
prompt photons, $N_{\gamma}$, i.e. neutral clusters in the EMC
having a time of flight consistent with a particle with $\beta=1$
coming from the interaction point. 
For signal counting we require $N_{\gamma}=2$, while for 
normalization purposes we require $N_{\gamma}=4$.
A tight constraint on $\beta$ is imposed to reduce the effect
of event losses due to accidental clusters from {\em dafbkg}.
Moreover, to improve the rejection of
the main background, $K_S \to 2 \pi^0$ with two lost photons,
we accept clusters with energy above 7 MeV and produced in a large 
angular acceptance, $|cos(\theta)|<0.93$. 
After this acceptance selection, the 
distribution of the lost photons for the background is
peaked in the forward direction, as shown by the simulation
in Fig.~\ref{QCALMC}.

\begin{figure}[!t]
\begin{center}
\epsfig{width=.9\textwidth,file=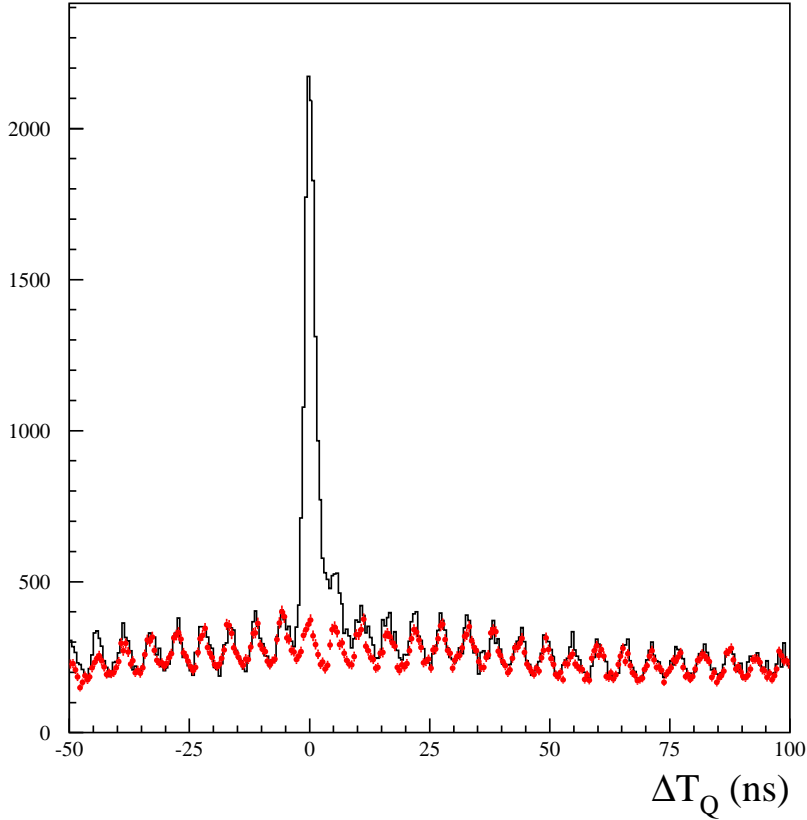}
\end{center} 
\caption{Inclusive distribution of the difference between the measured
arrival time and the expected time of flight of prompt photons in QCAL.
All hits for events tagged by a $K_L$-crash with $N_{\gamma}=2$ (black 
solid line) or with a reconstructed $K_S \to \pi^+\pi^-$ decay (points).}
\label{QCALT}
\end{figure}

To improve the background rejection we require a veto from the QCAL calorimeter.
This veto consists of rejecting events having at least
one hit in QCAL with energy above the pedestal and in time
with collisions. In Fig.~\ref{QCALT}, the data distribution of 
the difference, $\Delta T_{Q}$, between the reconstructed 
time of the QCAL hits, $T_{Q}$, and their expected time of 
flight, $TOF_{Q}$,  is shown for all tagged events with 
$N_{\gamma}=2$. 
The flat distribution of hits due to machine
background events shows clearly separated peaks bunched 
with the RF period. 
The sharp in-time peak observed is instead due to
the lost photons from the process $K_S \to 2 \pi^0$ impinging
on QCAL. We veto all events in a time window, TW, 
defined as $|\Delta T_{Q}| < 5$~ns.
The QCAL veto successfully rejects  $\sim 70$~\% of 
the background while retaining a very high efficiency ($\sim 99.96$ \%)  
on the signal.

Since we have not simulated the {\em dafbkg} events in QCAL ,
when applying the veto on  data we should correct 
for the losses due to the accidental coincidence of
these background hits in the time window used. 
A data-calibrated correction, $C_{Q} = 1-P^{TW}_{Q}$,
has been developed by determining $P^{TW}_{Q}$ i.e. 
the probability to find a spurious hit in TW.
To estimate it  we use two out-of-time 
windows one before, {\em early}, 
and one after, {\em late}, the collision time.
The average value of the probability in these two
control windows provides a first evaluation
of the correction. Furthermore, to assign a systematic
error to this determination, we have measured
these losses also in a control sample
with a "$K_L$-crash" and a well reconstructed
$K_S \to \pi^+\pi^-$ decay. This last sample 
does not have any photons impinging on
QCAL, thus allowing us to calculate directly the
losses in TW. The $\Delta T_{Q}$ distribution 
for these events is almost flat as shown
by the overlapped distribution (points) in Fig.~\ref{QCALT}.
We finally determine this probability to be:
$P^{TW}_{Q}=(3.51 \pm 0.04_{stat} \pm 0.26_{syst.}) \%$.

\begin{figure}[!t]
\begin{center}
\begin{tabular}{cc}
\hspace{-1cm}
\epsfig{width=.55\textwidth,file=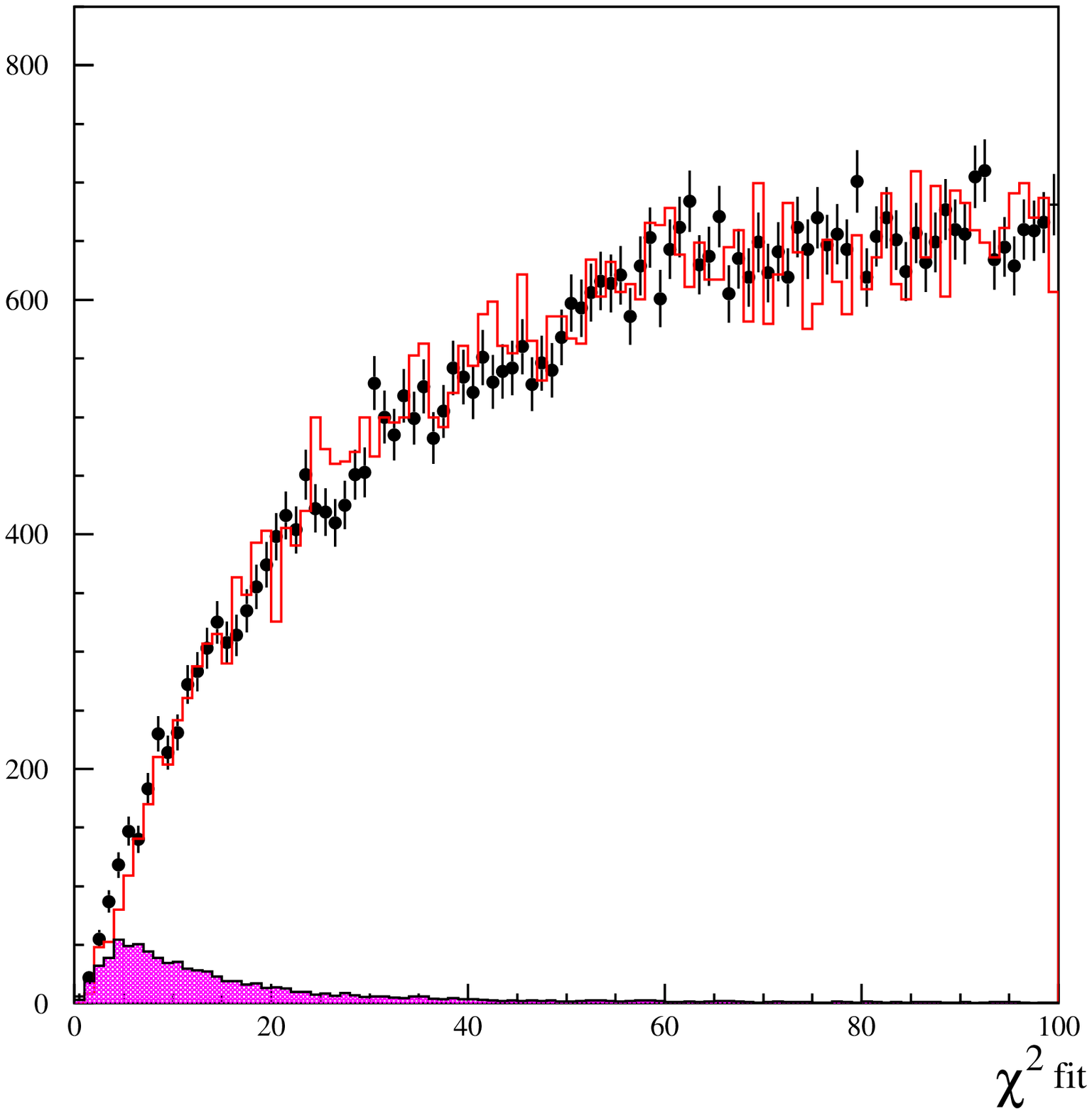} &
\hspace{-1cm}
\epsfig{width=.55\textwidth,file=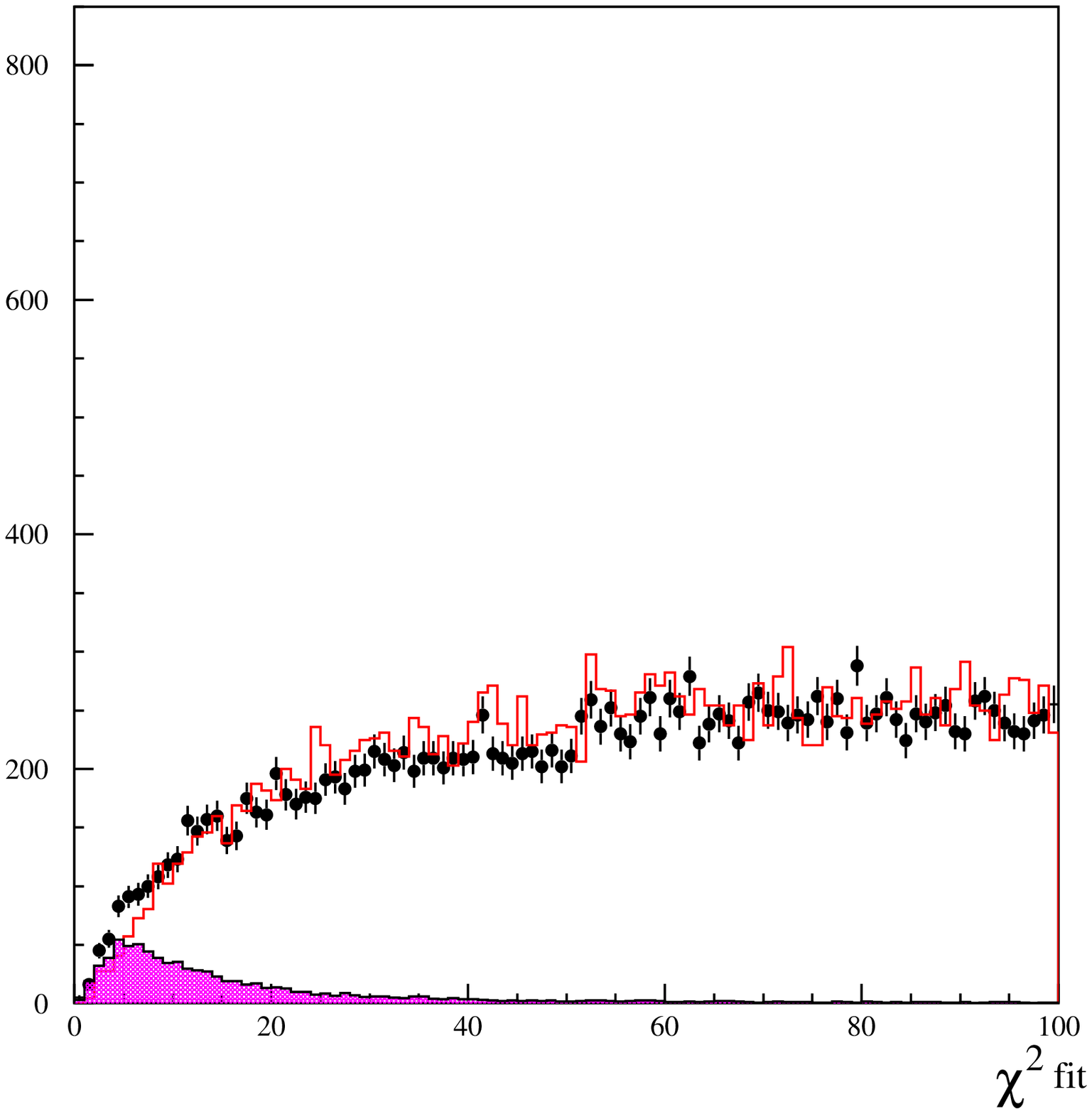} \\
\end{tabular}
\end{center} 
\caption{Distributions of $\chi^2_{FIT}$ for all events after tagging
and $N_{\gamma}=2$ requirement: before (left) and after (right)
the application of the QCAL Veto.} 
\label{chi2}
\end{figure}

At the end of the acceptance and QCAL veto selection, we count 
157 $\times 10^3$ events in data.
 
\begin{figure}[!t]
\begin{center}
\epsfig{width=.9\textwidth,file=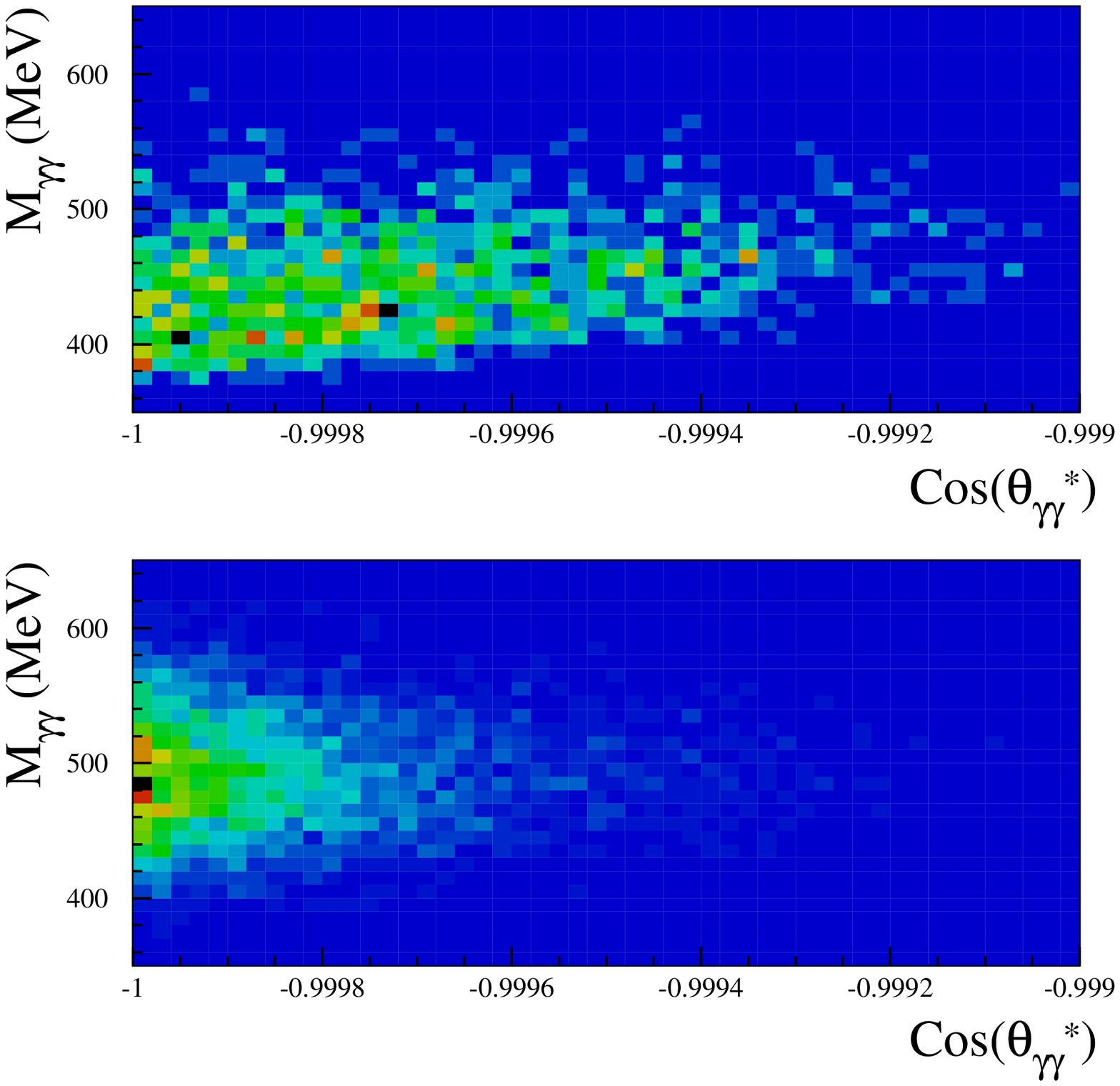}
\end{center} 
\caption{Scatter plot of $M_{\gamma \gamma}$ vs 
$cos(\theta^{*}_{\gamma\gamma})$ for simulated events:
background (top) and signal (bottom). }
\label{scatter_MC}
\end{figure}

\begin{figure}[!t]
\begin{center}
\epsfig{width=.7\textwidth,file=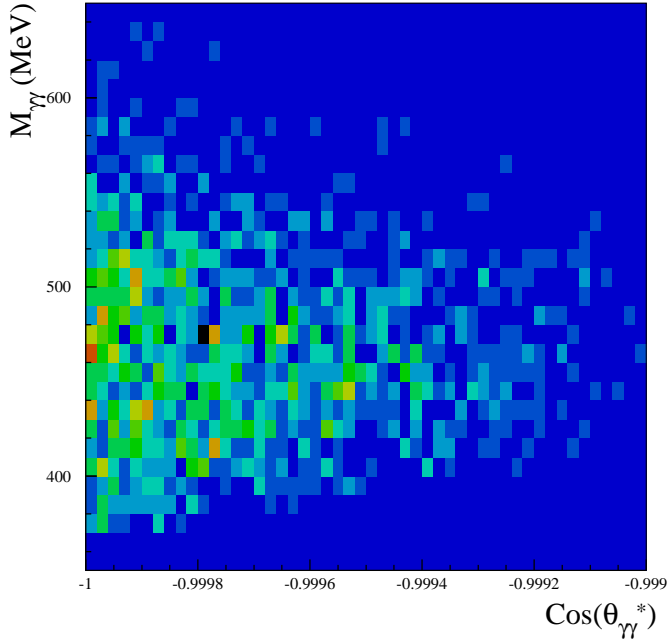}
\end{center} 
\caption{
Scatter plot of $M_{\gamma \gamma}$ vs 
$cos(\theta^{*}_{\gamma\gamma})$ for data after 
acceptance selection, QCAL veto and $\chi^2_{FIT}$ cut.}
\label{scatter_DATA}
\end{figure}

\begin{figure}[!t]
\begin{center}
\epsfig{width=1.0\textwidth,file=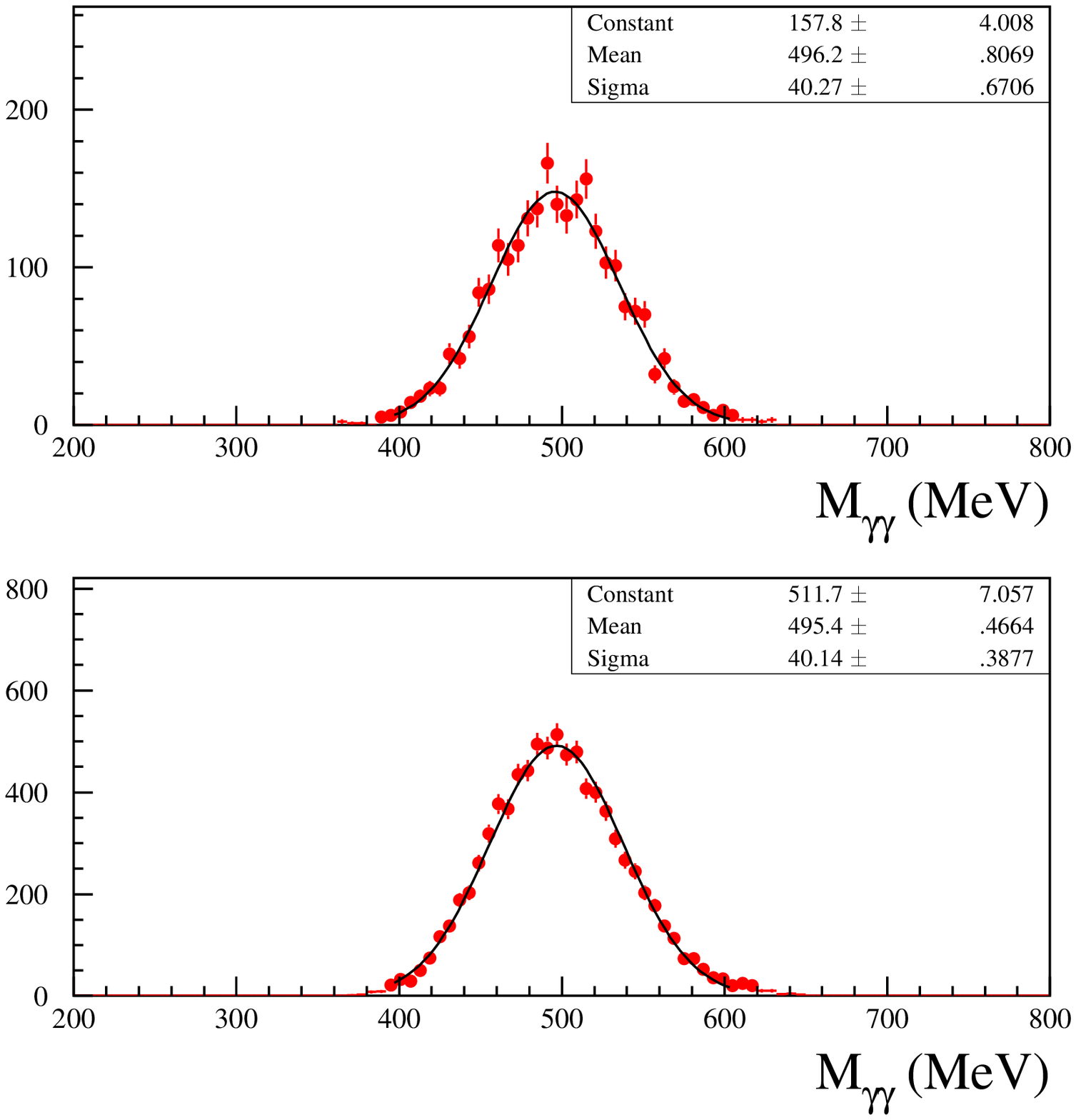} \\
\end{center} 
\caption{Distribution of the invariant mass of two photons, 
$M_{\gamma\gamma}$ (MeV), for the $K_L \to \gamma \gamma$ 
control sample with a neutral vertex in the fiducial region 
below the DC: data (top), MC samples (bottom). }
\label{energyscale}
\end{figure}


\subsection{Kinematic fitting and  event counting}
%
To improve the signal over background ratio (S/B), we apply
a kinematic fit procedure which imposes the
conservation of the $\phi$ 4-momentum  at the origin,
the $K_S$ mass and  $\beta=1$ for each photon (Ndof$=$7). The fit
uses the knowledge of the $K_L$ 4-momentum provided by 
the "$K_L$-crash" position, W and $\overline{P_{\phi}}$.
In Figs.~\ref{chi2}.left(.right) the distribution of the $\chi^2$
of this procedure, $\chi^2_{FIT}$, is shown for data
and MC after acceptance selection without (with) the application of
the  QCAL veto. A  large amount of the background is  distributed at 
high $\chi^2_{FIT}$  values while the signal is contained in the 
low $\chi^2_{FIT}$ region. 
In the following, we retain events by cutting at $\chi^2_{FIT}<20$.  
The Monte Carlo estimates that the  signal (background) efficiency 
of this cut is 63.3 \% (0.5\%).  The S/B greatly improves, from 1/500 
to 1/3.

Two other variables with a high discriminating power
against background are the invariant mass of the two photons, 
$M_{\gamma\gamma}$, and the opening angle between the two 
photons in the  $\ks$ center of mass system, 
$\theta^{*}_{\gamma\gamma}$. 
Since the kinematic fit imposes the \ks\, \kl\ direction 
and the $K_S$ mass, we use the reconstructed variables before 
fit constraining. In Fig.~\ref{scatter_MC}.top(bottom) the 2D-plot of
$M_{\gamma\gamma}$ as a function of $cos(\theta^{*}_{\gamma\gamma})$ 
is shown for background (signal) events.
In Fig.~\ref{scatter_DATA}, the distribution of the same
scatter-plot for data is also shown.

Before fitting the data with MC shapes, we have tested
our simulation ability of reproducing the signal by 
comparing with data a control sample of $K_L \to \gamma \gamma$ 
decaying near the beam pipe and tagged by $K_S \to \pi^+\pi^-$ events.
Around 200 pb$^{-1}$ of data and 450 pb$^{-1}$ of Monte Carlo
have been used. A kinematic fit-procedure, similar 
to the one of the $K_S \to \gamma\gamma$ sample, has been 
applied. The background  is reduced to a negligible quantity 
by retaining only the events with $\chi^2_{FIT}({\rm KL})<20$. A gaussian fit 
to the  $M_{\gamma\gamma}$  distribution of this control sample 
provides a central value of (496.2 $\pm$ 0.8) MeV in data
and of (488.7 $\pm$ 0.5) MeV in Monte Carlo. This corresponds to an
average energy-scale shift of $\sim$ 1\% in the simulation.
This motivated an in-depth data-MC comparison of energy
response and resolution as a function of the incident photon energy.
This study  has been carried out by looking at the energy pulls 
of the kinematic fit with a sample of $\sim 80 $ pb$^{-1}$ of
$K_S \to 2 \pi^0$   spread out over the entire data taking period. 
An ad-hoc correction has been applied to better calibrate
the simulation versus the data. After applying this correction, 
the comparison between data and MC for the $K_L \to \gamma \gamma$ 
control sample is improved as shown by the fit to the $M_{\gamma\gamma}$
distributions shown in Fig.~\ref{energyscale}.
This control sample has been also used  to make a data-MC
comparison for the $\chi^2_{FIT}({\rm KL})$ distribution. 
A good agreement  is also observed for this variable.

\begin{figure}[!t]
\begin{center}
\epsfig{width=1.0\textwidth,file=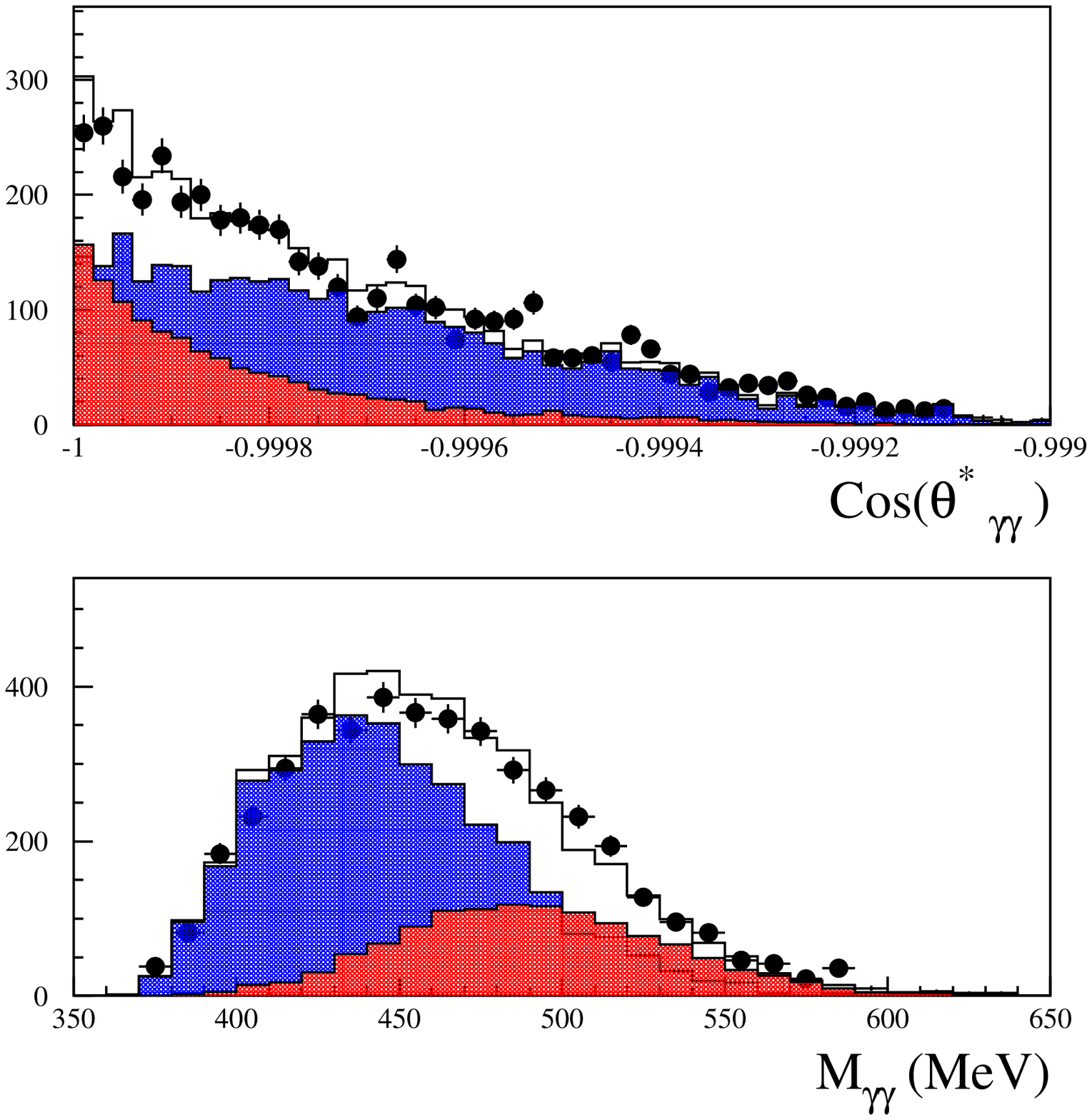}
\caption{Distributions of data (black points) after tagging,
acceptance selection and $\chi^2_{FIT}$ cut : 
cos$(\theta^{*}_{\gamma \gamma})$ (top), 
$M_{\gamma\gamma}$ (bottom). The superimposed colored
distributions are signal and background MC shapes.
Solid line is the sum of the MC shapes after fitting.}
\label{2dresult}
\end{center} 
\end{figure}

\begin{figure}[!t]
\begin{center}
\begin{tabular}{cc}
\hspace{-1cm}
\epsfig{width=.55\textwidth,file=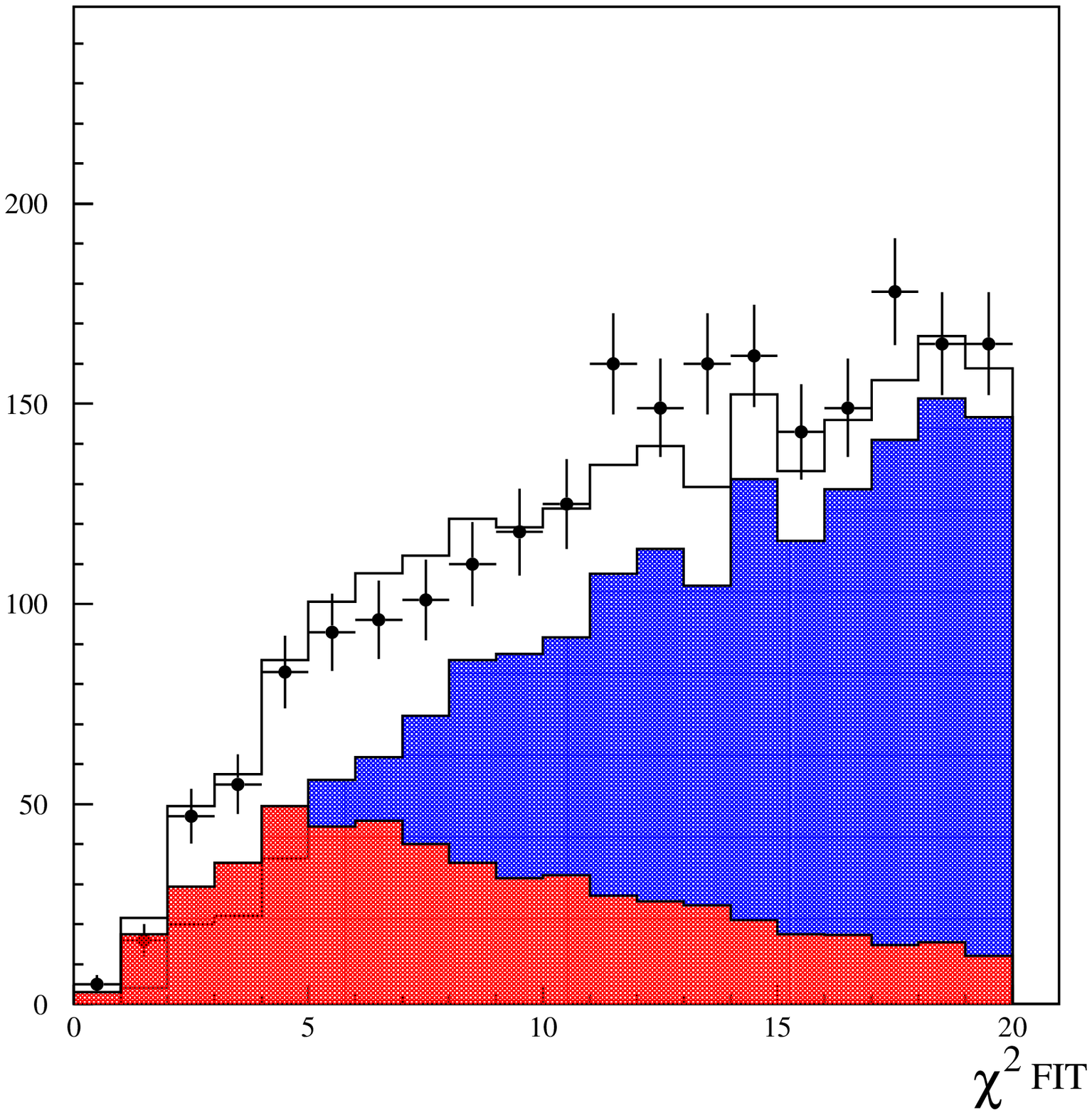} &
\hspace{-1cm}
\epsfig{width=.55\textwidth,file=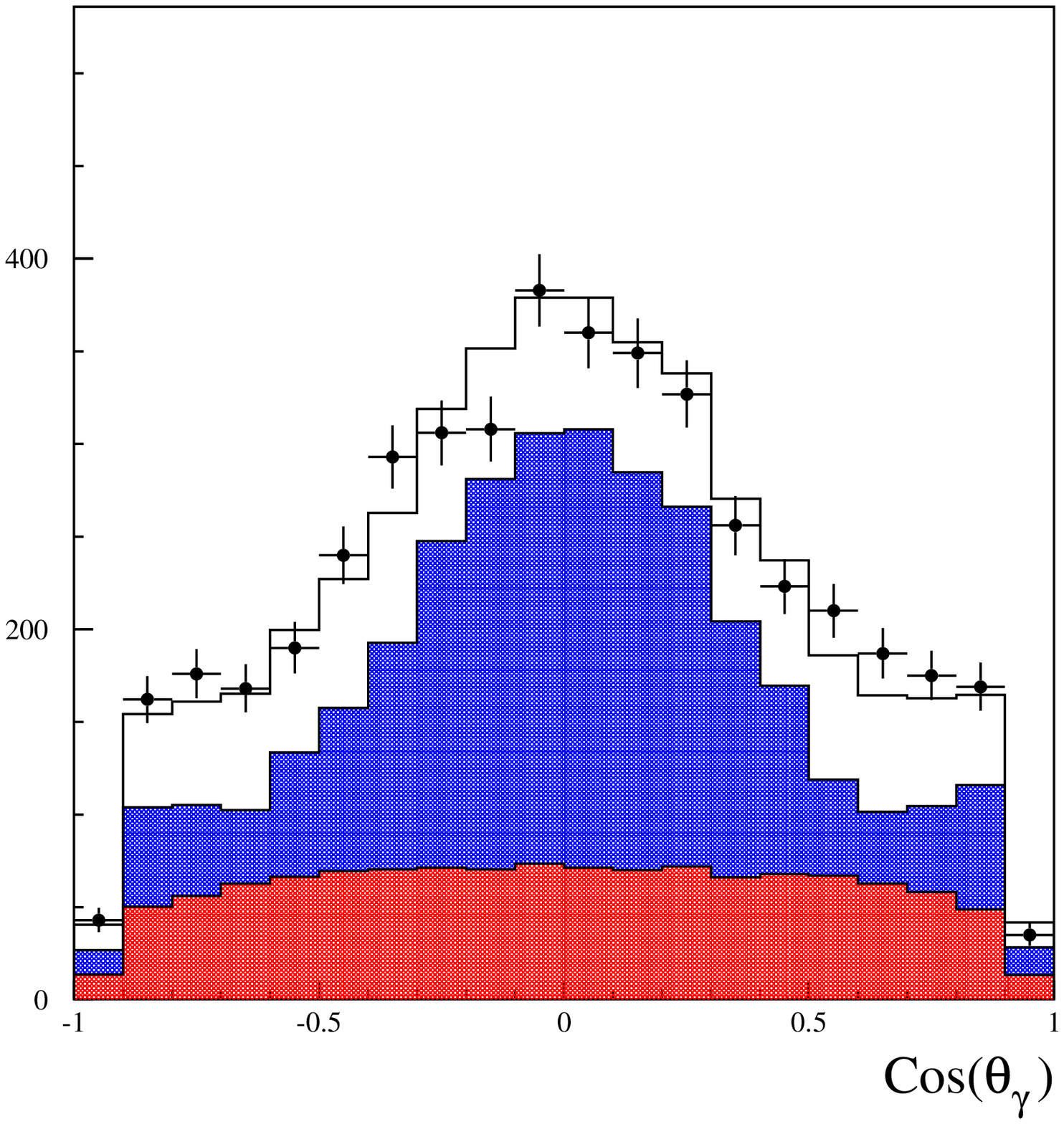} \\
\end{tabular}
\end{center} 
\caption{  
Distributions of data (black points) after tagging,
acceptance selection and $\chi^2_{FIT}$ cut : 
$\chi^2_{FIT}$ (left) and inclusive cos($\theta_{\gamma}$) 
for the two photons in the event (right). The superimposed colored
distributions are signal and background MC shapes.
Solid line is the sum of MC shapes after fitting.}
\label{control_plots}
\end{figure}

For the $K_S$-tagged events, a binned maximum-likelihood fit to 
the 2D $cos(\theta^{*}_{\gamma\gamma})$, $M_{\gamma\gamma}$ 
distribution on data is performed relying on
the MC signal and background shapes. 
The likelihood function properly takes into account data and 
MC statistics of the distributions.
The resulting $\chi^2/{\rm Ndof}$ of the fit is 1.2. The 
quality of the fit is shown in Fig.~\ref{2dresult} by comparing
with data the simulated shapes for mass and angular projections 
as weighed by the fit procedure.
As expected,  the  $cos(\theta^{*}_{\gamma\gamma})$ distribution
shows a signal shape which is  much  more peaked than the
background to $-1$. The $M_{\gamma\gamma}$ distribution 
shows a well identified gaussian shape around the $\ks$ mass for 
the signal, while  the background has an asymmetric shape peaked 
at lower mass values. We count $N(2\gamma)=600 \pm 35$ 
signal events out of  2280 events in the 2D plot.

At the end of the analysis chain and as an independent check 
of the quality of the fit weights found, we show in 
Fig.~\ref{control_plots}.left the $\chi^2_{FIT}$ distribution for data 
and MC as weighed by the scalar factors previously 
determined. A similar comparison is done also for the inclusive angular 
photon distribution (see Fig.~\ref{control_plots}.right).
The latter distribution clearly indicates the need for a flat 
angular dependence as expected by the uniform decay of a spin 
0 particle in two photons.

\section{Branching ratio evaluation and systematics}
The branching ratio is evaluated with respect to the
BR($K_S \to 2 \pi^0)$ by counting the $K_L$-crash tagged 
events with $N_{\gamma}=4$ in the same sample as follows:
\begin{equation}
BR(K_S \to 2\gamma) = \frac{N(2\gamma)}{N(2\pi^0)} \times
\frac{\varepsilon_{TOT}(2\pi^0|tag)}{\varepsilon_{TOT}(2\gamma|tag)}
\times BR(K_S \to 2\pi^0),
\end{equation}
where the total efficiencies have been evaluated by MC after
$K_S$ tagging. We have assumed that the ratio of trigger, 
event classification and tagging efficiencies between the 
two decays is one. The signal total efficiency  is the product 
of the efficiencies for the acceptance selection, the QCAL cut
and the $\chi^2$ cut. Each single one has been evaluated as 
conditioned efficiency.

The acceptance selection efficiency for the signal, after tagging,
 is
\begin{equation}
 \varepsilon_{sel}(2\gamma)= (83.2 \pm 0.2_{stat}\pm 0.1_{syst}) \%,
\end{equation}
this large efficiency is
due to the angular coverage of the calorimeter, 
the low energy threshold used and the flat angular 
distribution of the decay products.  The systematic 
error assigned to this efficiency has been found 
by varying the data-MC correction curves of the  cluster
reconstruction efficiency.
The efficiency of the QCAL cut, after tagging and acceptance, 
is found by MC to be $\sim$ 99.96 \% . However, on data 
we have to  apply  the corrections due to accidental losses described in 
sec.~\ref{accsel} to obtain:
\begin{equation}
\varepsilon_{Q} = \varepsilon_{Q}^{MC} \times C_{Q}
= (96.45\pm 0.04_{stat.} \pm 0.26_{syst.}) \%.
\end{equation}
The MC efficiency of the applied $\chi^2_{FIT}$ cut 
is $\varepsilon_{\chi^2} = (63.3 \pm 0.7)\%.$ 
To evaluate a systematic error due to the data-MC difference
in the $\chi^2_{FIT}$ scale, we have looked at
the $K_L \to \gamma \gamma$ control sample by requiring
loose $\chi^2_{FIT}$(KL) and angular cuts. A few percent contamination
exists. By building  $\chi^2_{FIT}$ cumulative distributions for 
data and MC and  calculating  their ratio at the applied cut value,  
a systematic error, $\Delta \varepsilon_{\chi^2}/\varepsilon_{\chi^2}$ 
of -0.4\% is assigned to this effect.

\begin{table}[!t]
\begin{center}
\begin{tabular}{|c|c|c|}
\hline
Source & +$\Delta BR/BR$ (\%) & 
-$\Delta BR/BR$ (\%) \\
\hline
Signal acceptance & 0.12 & 0.12 \\
QCAL losses & 0.02 & 0.26 \\
$\chi ^2$ scale & --  & 0.41 \\
\hline
Normalization sample & 0.15 & 0.15 \\
\hline
QCAL TW change  & 0.88 & 0.44 \\
$\chi^2$ change & 0.44 & 0.44 \\
2D-Fit binning  & 0.88 & 0.44 \\
 MC Energy scale & -- & 1.32 \\
\hline
\hline
Total & 1.33  &  1.61 \\
\hline
\end{tabular}
\end{center}
\caption{Contributions to the total systematic error on the
BR. The first three contributions have been evaluated 
directly as systematics related to the signal efficiency 
for a given cut. The fourth contribution regards the 
systematics on efficiency for the normalization sample.
The last four contributions have been evaluated by repeating
the BR measurement while varying analysis conditions or cuts.}
\label{TABSYS}
\end{table}

For the normalization  we have counted $K_S \to 2\pi^0$ tagged
events with $N_{\gamma}=4$. An efficiency of
\begin{equation} 
\varepsilon_{sel}(4\gamma) = (65.0 \pm 0.2_{stat} \pm {0.1}_{syst}) \% 
\end{equation}
is found by Monte Carlo. The systematics has been evaluated,
as done for the signal, by varying the data-MC correction curves 
of the  cluster reconstruction efficiency. After correcting
for $\varepsilon_{sel}(4\gamma)$, a total number of
$(159.8 \pm 0.5) \times 10^6$  $K_S \to 2\pi^0$
tagged events is obtained. Another systematic uncertainty 
related to accidental overlap of {\em dafbkg}
clusters, shower fragmentation and merging of nearby clusters 
has been  evaluated by repeating the measurement
in an inclusive way and counting tagged events with 3, 4 and 5 
photons. We get an efficiency corrected counting of $159.5 \times 10^6$
events which agrees at $\sim 2.5$  per mil level with the number
obtained with the exclusive counting.

\begin{figure}[!t]
\begin{center}
\begin{tabular}{cc}
\hspace{-1cm}
\epsfig{width=0.55\textwidth,file=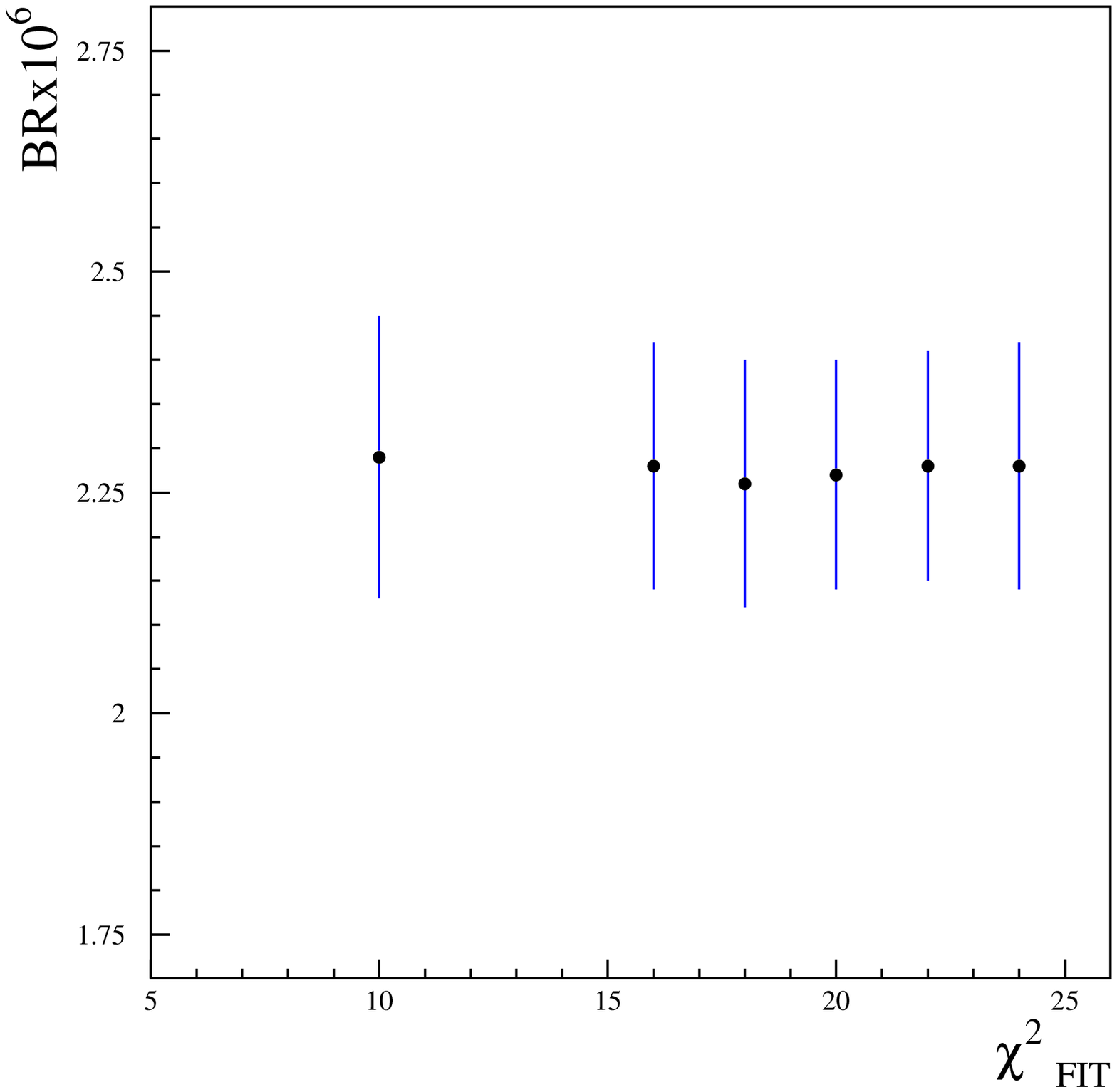}  &
\hspace{-1cm}
\epsfig{width=0.55\textwidth,file=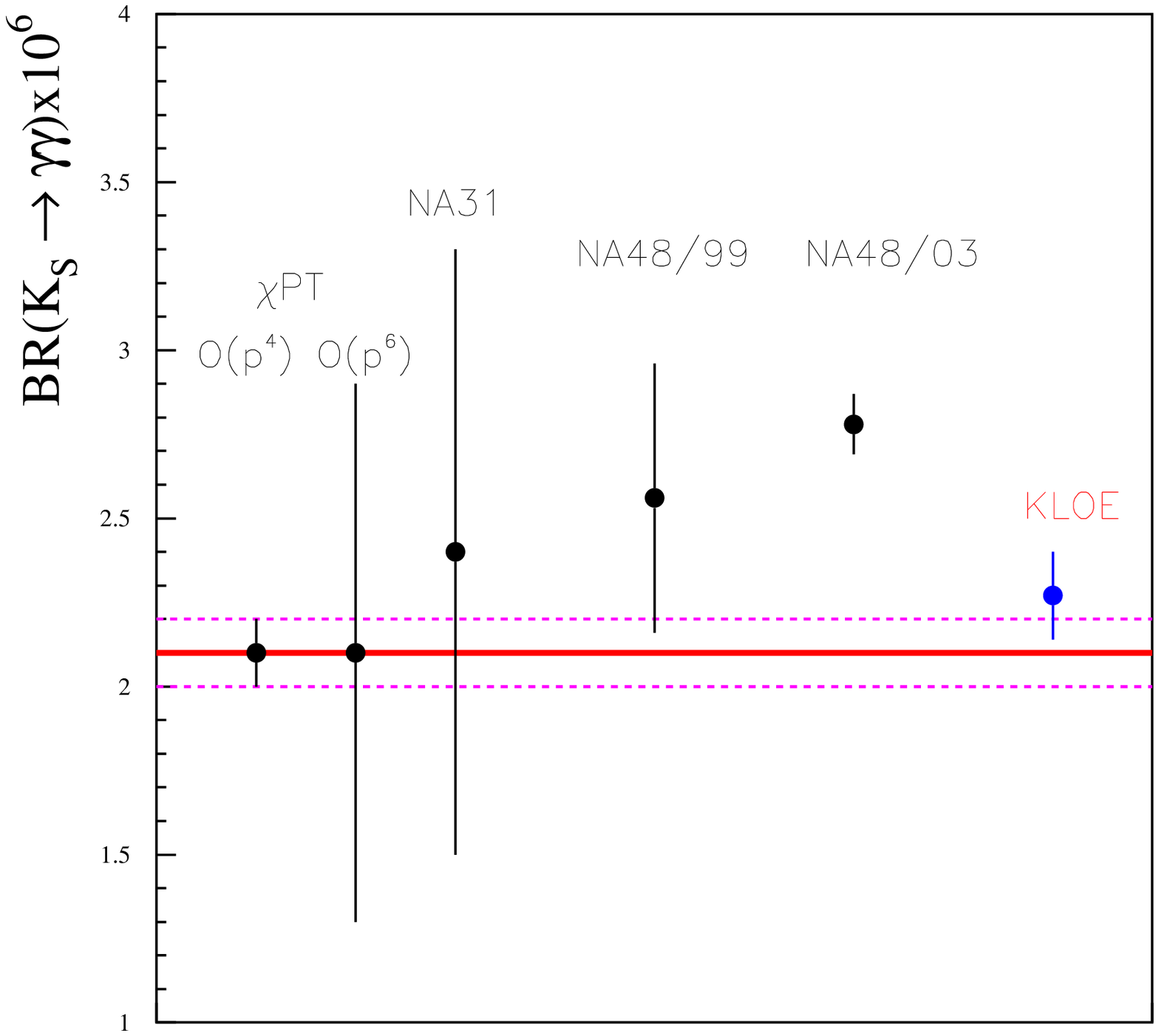} \\
\end{tabular}
\caption{Variation of BR result as a function of the
applied $\chi^2_{FIT}$ cut (left). Comparison of our
measurement of BR$(K_S \to \gamma \gamma)$ with the other 
existing measurements and $O(P^4)$ $\chi$PT predictions (right). }
\label{finalres} 
\end{center}
\end{figure}

For the BR($K_S\to\gamma\gamma$) we use the
latest PDG~\cite{pdg06} value of BR($K_S \to 2 \pi^0$) which is 
$(30.69 \pm 0.05)$\%. The systematics connected to the counting 
has been evaluated by repeating the analysis in different ways. 
We first tested the stability of the branching ratio when 
modifying the width of the time window used for the QCAL veto 
or the applied value of the $\chi^2_{FIT}$ cut. In Fig.~\ref{finalres}.left 
the BR changes as a function of the applied $\chi^2_{FIT}$
cut is shown. We then repeated the fit by applying the residual 
energy scale shift of +0.4\% to the MC distributions and varied 
the bin size used in the 2D plot for fitting. In all 
cases, the maximum variation of the BR is used as systematic error
and shown in Tab.~\ref{TABSYS}. The sum in quadrature of all entries
in the table is used as total systematic error.
We obtain:
\begin{equation}
BR(K_S \to \gamma \gamma) = (2.27 \pm 0.13(stat.) ^{+0.03}_{-0.04} (syst.))\times 10^{-6}.
\end{equation}

\section{Conclusion}

With a sample of 1.62 fb$^{-1}$ of $e^+e^-$ collisions at $\sqrt{s} \sim M_{\phi}$
collected with KLOE at DA$\Phi$NE,
we have measured the $BR(K_S\to \gamma\gamma)$ with a 5.6\% statistical 
uncertainty and a  $\sim 1.5$ \% systematic error. We obtain
a BR result which deviates by 2.9 $\sigma$'s  from the previous
best precise determination, as shown in Fig.~\ref{finalres}.right.
Our measurement is also consistent, within errors, with $O(P^4)$ 
$\chi PT$   predictions.


\begin{thebibliography}{99}
\bibitem{ambrosio}G.~D'Ambrosio, D.~Espriu,
Phys. Lett. {\bf B175} (1986).
\bibitem{na31}G.~D.~Barr, {\it at al.},
Phys. Lett. {\bf B493} (1995).
\bibitem{na48}J.~R.~Batley, {\it at al.},
Phys. Lett. {\bf B551} (2003).
\bibitem{kloe1} KLOE collaboration, M.~Adinolfi {\it et al.}, 
Nucl. Inst. Meth. A 488 (2002), 51. 
\bibitem{kloe2} KLOE collaboration, M.~Adinolfi {\it et al.}, 
Nucl. Inst. Meth. A 482 (2002), 364.
\bibitem{kloe3} KLOE collaboration, M.~Adinolfi {\it et al.}, 
Nucl. Inst. Meth. A 483 (2002), 649.
\bibitem{kloe4} KLOE collaboration, M.~Adinolfi {\it et al.}, 
Nucl. Inst. Meth. A 492 (2002), 134.
\bibitem{dafne} S.~Guiducci, in: P.~Lucas, S.~Weber (Eds.),
Proceedings of the 2001 Particle Accelerator Conference, Chicago, Il.,
USA, 2001.
\bibitem{NIM} KLOE collaboration, F.~Ambrosino {\it et al.},
Nucl. Inst. Meth. A 534 (2004), 403.
\bibitem{pdg06} W.M.Yao {\it et al.}, Journal of Physics G 33, (2006), 1.
\end{thebibliography}
\end{document}